\documentclass{ws-ijmpa}

\begin{document}

\markboth{V\'{\i}ctor M. Villalba and Clara Rojas} {Bound states of
the Klein-Gordon equation}

%
\catchline{}{}{}{}{}
%

\title{BOUND STATES OF THE KLEIN-GORDON EQUATION IN THE PRESENCE OF SHORT RANGE POTENTIALS
}

\author{\footnotesize V\'ICTOR M. VILLALBA\footnote{e-mail
villalba@ivic.ve} \hspace{0.1cm}
 and  CLARA ROJAS}

\address{Centro de F\'{\i}sica, Apdo 21827, Caracas 1020A. Venezuela }

\maketitle

\pub{Received (Day Month Year)}{Revised (Day Month Year)}

\begin{abstract}
We solve the Klein-Gordon equation in the presence of a spatially
one-dimensional cusp potential. The bound state solutions are
derived and the antiparticle bound state is discussed.
\end{abstract}

\section{Introduction}

Since the appearance of the pioneering paper of Snyder and
Weinberg\cite{Snyder}, many articles have been published on the
problem of meson fields in the presence of strong electric and
gravitational fields. In 1940 Schiff, Snyder and
Weinberg\cite{Schiff} carried out one of the earliest investigations
of the solution of the Klein-Gordon equation with a strong external
potential. They solved the problem of the square well potential and
discovered that there is a critical point $V_{cr}$ where the bound
antiparticle mode appears to coalesce with the bound particle.
Popov\cite{Popov,Rafelski} showed that this phenomenon is a
particular effect characteristic of short range potentials and,
consequently it should not be expected to be observed when one deals
with Coulomb interactions\cite{Rafelski,Bawin2,Klein,Klein2} The
asymptotic limit of the Schiff-Snyder effect, for infinite walls,
has been studied by Fulling\cite{Fulling}. In $1979$, Bawin and
Lavine\cite{Bawin} demonstrated that the antiparticle $p$-wave bound
state arises for some conditions on the potential parameters,
showing in this way that, for short range potentials, the presence
of the Schiff-Snyder effect strongly depends on the angular momentum
contribution.

The presence of strong fields introduces quantum phenomena, such as
supercriticallity and spontaneous pair production, that cannot be
described using perturbative techniques. The discussion of
overcritical behavior of bosons requires a full understanding of the
single particle spectrum, and consequently of the exact solutions to
the Klein-Gordon equation. Overcritical behavior associated with
scalar particles in strong short range potentials could be of
interest in understanding quantum effects such as superradiance\cite
{Fulling,Zeldovich} or particle  production in the vicinity of black
holes \cite{Fulling,Birrell}.

In the present article, we solve the Klein-Gordon equation for a
Woods-Saxon and cusp potentials. The interest in computing bound
states and spontaneous pair creation processes in such potentials
lies in the fact that they possess properties that could permit us
to determine how the shape of the potential affects the pair
creation mechanism. We show that the antiparticle bound states arise
also for the Woods-Saxon potential well, which is a smoothed out
form of the square well. We also show that a cusp potential well
supports antiparticle bound states.

The article is structured as follows: Sec. 2 is devoted to discuss
the Klein-Gordon equation. In Sec. 3 we solve the Klein-Gordon
equation in the presence of a one-dimensional Woods-Saxon potential
well. We derive the equation governing the eigenvalues corresponding
to the bound states and compute the bound states. In Sec. 4 we
compute the bound states of the Klein-Gordon equation in the
presence of an one-dimensional cusp potential well. We also show the
dependence of supercritical states on the strength and shape of the
potential. Finally, in Sec. 5, we briefly summarize our results.

\section{The Klein-Gordon equation}

The Klein-Gordon equation minimally coupled to a vector potential
$A^{\mu }$ can be written as

\begin{equation}  \label{uno}
\eta ^{\mu \nu }(\partial _{\mu }+ieA_{\mu })(\partial _{\nu
}+ieA_{\nu })\phi +m^{2}\phi =0,
\end{equation}

\noindent where the metric $\eta ^{\mu \nu }$ has signature $-2$,
and we have set $\hbar =c=1.$ Choosing to work with a vector
potential of the form,

\begin{equation}
eA^{0}=V(\mathbf{r}),\ \vec{A}=0,
\end{equation}

\noindent Eq. (\ref{uno}) can be written as:

\begin{equation}  \label{dos}
\left(\frac{\partial }{\partial t}+iV(\mathbf{r})\right)^{2}\phi
-\nabla ^{2}\phi +m^{2}\phi =0.
\end{equation}
Eq. (\ref{dos}) can be completely separated in spherical coordinates
when one works with radial potentials\cite{Greiner,Greiner2}.
Analogously, we have that in Cartesian coordinates, the Klein-Gordon
equation can be reduced to a second order ordinary differential
equation when the potential depends only on a single space variable
$x.$  In this case Eq. (\ref{dos}) takes the simple form:

\begin{equation}
\frac{d^{2}\phi (x)}{dx^{2}}+\left[ \left( E-V(x)\right)
^{2}-m^{2}-p_{\bot }^{2}\right] \phi (x)=0,  \label{0}
\end{equation}
where, since the potential $V(x)$ does not depend on time, we have
separated variables in the form

\begin{equation}
\phi (x,t)=\phi (x)\exp (-iEt+ip_{y}y+ip_{z}z),
\end{equation}

\noindent and $p_{\bot }^{2}=p_{y}^{2}+p_{z}^{2}$.

Eq. (\ref{0}) has the same structure than the one obtained when one
solves the Klein-Gordon equation for $s$ states, in spherical
coordinates, for radial potentials\cite{Bawin2,Greiner}, therefore
the results reported in this article are also valid for $s$-waves in
radial Woods-Saxon and cusp potentials. In the next section we solve
Eq. (\ref{0}) in the presence of a Woods-Saxon potential well.

\section{The Woods-Saxon well}

\begin{figure}
\vspace{1cm} \centerline{\psfig{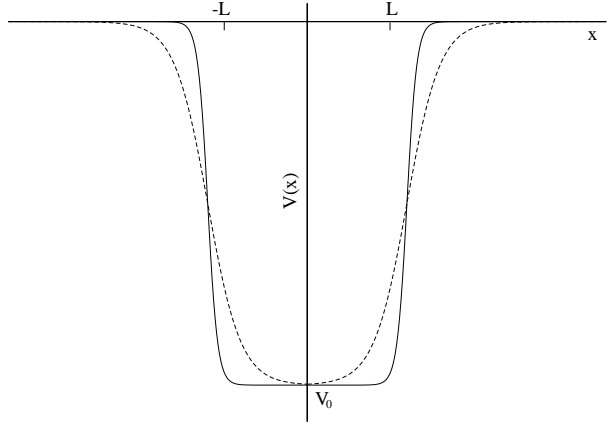}}
\vspace*{8pt} \caption{The Woods-Saxon potential well for $L=2$ with
$a=10$ (solid line) and $a=3$ (dotted line).} \label{fig:Fig1}
\end{figure}

The Woods-Saxon potential is defined as\cite{Kennedy}

\begin{equation}
V(x)=-V_{0}\left[ \frac{\Theta (-x)}{1+e^{-a(x+L)}}+\frac{\Theta (x)}{%
1+e^{a(x-L)}}\right] ,  \label{ws1}
\end{equation}
where $V_{0}$ is real and positive for a well potential; $a>0$ and
$L>0$ are real and positive. $\Theta (x)$ is the Heaviside step
function. The parameter $a$ defines the shape of the barrier or
well. The form of the Woods-Saxon potential is shown in Fig.
\ref{fig:Fig1}.
In order to solve the one-dimensional Klein-Gordon equation (\ref{0}), with $%
p_{\perp }=0$, \ in the presence of the Woods-Saxon potential
(\ref{ws1}) we split the solutions into two regions. Let us consider
bound states solutions for $x<0$. In this case we need to solve the
ordinary differential equation

\begin{equation}  \label{ws2}
\frac{d^{2}\phi _{L}(x)}{dx^{2}}+\left[ \left( E+\frac{V_{0}}{1+e^{-a(x+L)}}%
\right) ^{2}-1\right] \phi _{L}(x)=0,
\end{equation}

\noindent where in Eq. (\ref{ws2}) and thereafter we have equated the mass $%
m $ to unity. Since the potential (\ref{ws1}) depends only on one
space variable, the contribution of the transverse momentum in Eq.
(\ref{0}) appears via the introduction of an additive term that can
be included in an effective mass term $m_{eff}$ of the form
$m_{eff}=(m^2+p_{\bot }^{2})^{1/2}$. For small values of $p_{\bot
}$, the contribution of the transverse momentum does not introduce
qualitative changes to the energy spectrum.

On making the substitution $y^{-1}=1+e^{-a(x+L)}$, Eq. (\ref{ws2})
becomes

\begin{equation}  \label{ws3}
a^{2}y(1-y)\frac{d}{dy}\left[ y(1-y)\frac{d\phi _{L}(y)}{dy}\right]
+\left[ \left( E+V_{0}y\right) ^{2}-1\right] \phi _{L}(y)=0.
\end{equation}

Setting $\phi _{L}(y)=y^{\sigma }(1-y)^{\gamma }h(y)$ and
substituting it into Eq. (\ref{ws3}), we obtain the hypergeometric
equation

\begin{equation}  \label{ws4}
y(1-y)h^{\prime\prime}+[(1+2\sigma)-2(\sigma+\gamma+1)y]h^{\prime}-\left({%
\scriptstyle{\frac 1 2}} + \sigma + \gamma + \lambda\right)\left({%
\scriptstyle{\frac 1 2}} + \sigma + \gamma - \lambda\right)h=0,
\end{equation}
where the prime denotes a derivative with respect to $y$ and the
parameters $\sigma ,$ $\gamma $, and $\lambda $ are

\begin{equation}
\sigma =\frac{\sqrt{1-E^{2}}}{a},  \label{ws5}
\end{equation}

\begin{equation}
\gamma =\frac{\sqrt{1-(E+V_{0})^{2}}}{a}, \hspace{0.3cm} \lambda =\frac{%
\sqrt{a^{2}-4V_{0}^{2}}}{2a}.  \label{ws6}
\end{equation}
The general solution of Eq. (\ref{ws4}) can be expressed in terms of
Gauss hypergeometric functions \cite{Abra}:

\begin{eqnarray}
&&h(y)=a_{1}\hspace{0.05cm}_{2}F_{1}\left( \frac{1}{2}+\gamma
+\sigma -\lambda ,\frac{1}{2}+\gamma +\sigma +\lambda ,1+2\sigma
;y\right)  \nonumber
\label{ws8} \\
&&\quad \quad \hspace{0cm}+\hspace{0cm}a_{2}y^{-2\sigma }\hspace{0.05cm}%
_{2}F_{1}\left( \frac{1}{2}+\gamma -\sigma -\lambda
,\frac{1}{2}+\gamma -\sigma +\lambda ,1-2\sigma ;y\right) ,
\end{eqnarray}

\vspace{-0.4cm} \noindent so \vspace{-0.2cm}

\begin{eqnarray}
&&\phi _{L}(y)={a_{1}}y^{\sigma }(1-y)^{\gamma }\hspace{0.05cm}%
_{2}F_{1}\left( \frac{1}{2}+\gamma +\sigma -\lambda
,\frac{1}{2}+\gamma
+\sigma +\lambda ,1+2\sigma ;y\right)  \nonumber  \label{ws7} \\
&&\quad \quad \hspace{0.2cm}+\hspace{0.1cm}a_{2}y^{-\sigma }(1-y)^{\gamma }%
\hspace{0.05cm}_{2}F_{1}\left( \frac{1}{2}+\gamma -\sigma -\lambda ,\frac{1}{%
2}+\gamma -\sigma +\lambda ,1-2\sigma ;y\right) .
\end{eqnarray}

Now we consider the solution for $x>0$. We solve the differential
equation

\begin{equation}
\frac{d^{2}\phi _{R}(x)}{dx^{2}}+\left[ \left( E+\frac{V_{0}}{1+e^{a(x-L)}}%
\right) ^{2}-1\right] \phi _{R}(x)=0.  \label{ws9}
\end{equation}
After making the substitution $z^{-1}=1+e^{a(x-L)}$, Eq. (\ref{ws9})
can be written as:

\begin{equation}  \label{ws10}
a^2 z (1-z)\frac{d}{dz}\left[z(1-z)\frac{d\phi_R(z)}{dz}\right] +\left[%
\left(E+V_0z \right)^2-1 \right]\phi_R(z)=0.
\end{equation}
Introducing $\phi_R(z)=z^\sigma(1-z)^{-\gamma} g(z)$ and
substituting it into Eq. (\ref{ws10}), we obtain that $\phi_R(z)$
satisfies the hypergeometric equation:

\begin{equation}  \label{ws11}
z(1-z)g^{\prime\prime}+[(1+2\sigma)-2(\sigma-\gamma+1)z]g^{\prime}
\nonumber
\\
-\left({\scriptstyle{\frac 1 2}} + \sigma - \gamma + \lambda\right)\left({%
\scriptstyle{\frac 1 2}} + \sigma - \gamma - \lambda\right)g=0,
\end{equation}

\noindent where the prime denotes a derivative with respect to $z$.
The general solution of Eq. (\ref{ws11}) is \cite{Abra}:

\begin{eqnarray}
&&g(z)={b_{1}}_{2}F_{1}\left( \frac{1}{2}-\gamma +\sigma -\lambda ,\frac{1}{2%
}-\gamma +\sigma +\lambda ,1+2\sigma ;z\right)   \nonumber  \label{ws12} \\
&&\ \ \ \ \ \ \ \ +\hspace{0.1cm}b_{2}z^{-2\sigma }\hspace{0.05cm}%
_{2}F_{1}\left( \frac{1}{2}-\gamma -\sigma -\lambda
,\frac{1}{2}-\gamma -\sigma +\lambda ,1-2\sigma ;z\right) ,
\end{eqnarray}

\vspace{-0.4cm} \noindent so \vspace{-0.2cm}

\begin{eqnarray}
&&\phi _{R}(z)=b_{1}z^{\sigma }(1-z)^{-\gamma }\hspace{0.05cm}%
_{2}F_{1}\left( \frac{1}{2}-\gamma +\sigma -\lambda
,\frac{1}{2}-\gamma
+\sigma +\lambda ,1+2\sigma ;z\right)   \nonumber  \label{ws13} \\
&&\quad \ \ \ \ +\hspace{0.1cm}b_{2}z^{-\sigma }(1-z)^{-\gamma }\hspace{%
0.05cm}_{2}F_{1}\left( \frac{1}{2}-\gamma -\sigma -\lambda ,\frac{1}{2}%
-\gamma -\sigma +\lambda ,1-2\sigma ;z\right) .
\end{eqnarray}

As $x\rightarrow-\infty$ we have that  $y$ goes to zero,
Analogously, as $x\rightarrow \infty$, $z$ also goes to zero. We
choose the regular wave functions

\begin{eqnarray}
&&\phi _{L}(y)=a_{1}y^{\sigma }(1-y)^{\gamma
}\hspace{0.05cm}_{2}F_{1}\left( \frac{1}{2}+\gamma +\sigma -\lambda
,\frac{1}{2}+\gamma +\sigma +\lambda
,1+2\sigma ;y\right) .  \nonumber  \label{ws14} \\
&&\phi _{R}(z)=b_{1}z^{\sigma }(1-z)^{-\gamma }\hspace{0.05cm}%
_{2}F_{1}\left( \frac{1}{2}-\gamma +\sigma -\lambda
,\frac{1}{2}-\gamma +\sigma +\lambda ,1+2\sigma ;z\right) .
\end{eqnarray}

In order to find the energy eigenvalues, we impose that the right
and left wave functions and their first derivatives must be matched
at $x=0$. This condition leads to

\begin{eqnarray}
&&\frac{1}{1+2\sigma }\biggl[\scriptstyle{\left( \frac{1}{2}+\gamma
+\sigma
-\lambda \right) \left( \frac{1}{2}+\gamma +\sigma +\lambda \right) }\biggr.%
\times \frac{_{2}F_{1}\left( \frac{3}{2}+\gamma +\sigma -\lambda ,\frac{3}{2}%
+\gamma +\sigma +\lambda ,2+2\,\sigma ,({1+e^{-aL}})^{-1}\right) }{%
_{2}F_{1}\left( \frac{1}{2}+\gamma +\sigma -\lambda
,\frac{1}{2}+\gamma +\sigma +\lambda ,1+2\,\sigma
,({1+e^{-aL}})^{-1}\right) }  \nonumber
\label{ws15} \\
&&+\hspace{0.3cm}\scriptstyle{\left( \frac{1}{2}-\gamma +\sigma
-\lambda \right) \left( \frac{1}{2}-\gamma +\sigma +\lambda \right)
}\biggl.\ \times
\frac{_{2}F_{1}\left( \frac{3}{2}-\gamma +\sigma -\lambda ,\frac{3}{2}%
-\gamma +\sigma +\lambda ,2+2\,\sigma ,({1+e^{-aL}})^{-1}\right) }{%
_{2}F_{1}\left( \frac{1}{2}-\gamma +\sigma -\lambda
,\frac{1}{2}-\gamma +\sigma +\lambda ,1+2\,\sigma
,({1+e^{-aL}})^{-1}\right) }\biggr]  \nonumber
\\
&&\hspace{0cm}+\hspace{0.3cm}\scriptstyle{2\sigma \left(
1+e^{-aL}\right) \quad =\quad 0},
\end{eqnarray}
which is the eigenvalue condition for the energy E. Explicit
solutions of Eq. (\ref{ws15}), giving E in terms of $V_{0}$, can be
determined numerically. We consider the range $-1<E<1$ for the values of $E$%
. Some aspects of the dependence of the spectrum of bound states on
the
potential strength $V_{0}$ are shown in Figs. 2 and 3. At some value of $%
V_{0}$, a bound antiparticle state appears, it joins with the bound
particle state, they form a state with zero norm at $V_{0}=V_{cr}$
and then both vanish from the spectrum.

The normalization of the wave functions in Eq. (\ref{ws14}) is given
by

\begin{equation}  \label{norma}
N=2\int_{-\infty }^{\infty }dx[E-V(x)]\phi (x)^{\ast }\phi (x).
\end{equation}

The norm of the Klein-Gordon equation vanishes at $V_{cr}$, where
both possible solution $E^{(+)}$ and $E^{(-)}$ meet. Fig. 2. shows
that for $2.0900 < V_0 < 2.0908$ two states appear, one particle (solid line), 
and one antiparticle state (dashed line). In Fig. 3 the same
behavior is observed for $2.3462 < V_0 < 2.3463$. Particle bound states ($%
E^{(+)}$) and antiparticle bound states ($E^{(-)}$) correspond to $N>0$ and $%
N < 0$ respectively. For $N=0$ both solutions meet and have the same
energy. Antiparticle states appear in all the cases considered. For
$L=2$, we moved
the shape parameter $a$ from $1$ to $18$ and, for $L=1$, we considered $a=10$%
. Fig. 4 and Fig. 5 show the behavior of the turning point $(E)$
versus the potential parameters $a$ and $V_0$ respectively. Fig. 4
shows that as the value of $a$ increases, the energy value, for
which antiparticle states appear, increases. In Fig. 5 we observe
that, as the value of $V_0$ increases, the energy value for which
antiparticle states appear decreases. This behavior indicates that
square well potentials exhibit antiparticle bound states for values
of $E$ larger than for smoothed out potentials.

\begin{figure}[th]
\vspace{1cm} \centerline{\psfig{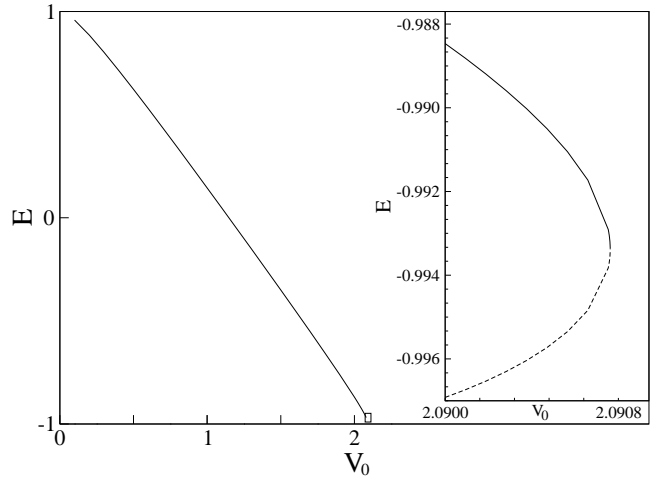}}
\vspace*{0pt} \caption{Energy of the lowest bound-state spectrum for
$L=2$, $a=10$. Inset is an enlargement of the critical area, showing
solid and dotted lines corresponding to positive and
negative norm state solutions respectively. The critical value for
$V_{0}=2.0908$ corresponds to $E=-0.993698$. Energy is given in units
of the rest energy $mc^{2}$} \label{fig:Fig2}
\end{figure}

\begin{figure}[th]
\vspace{1cm} \centerline{\psfig{file=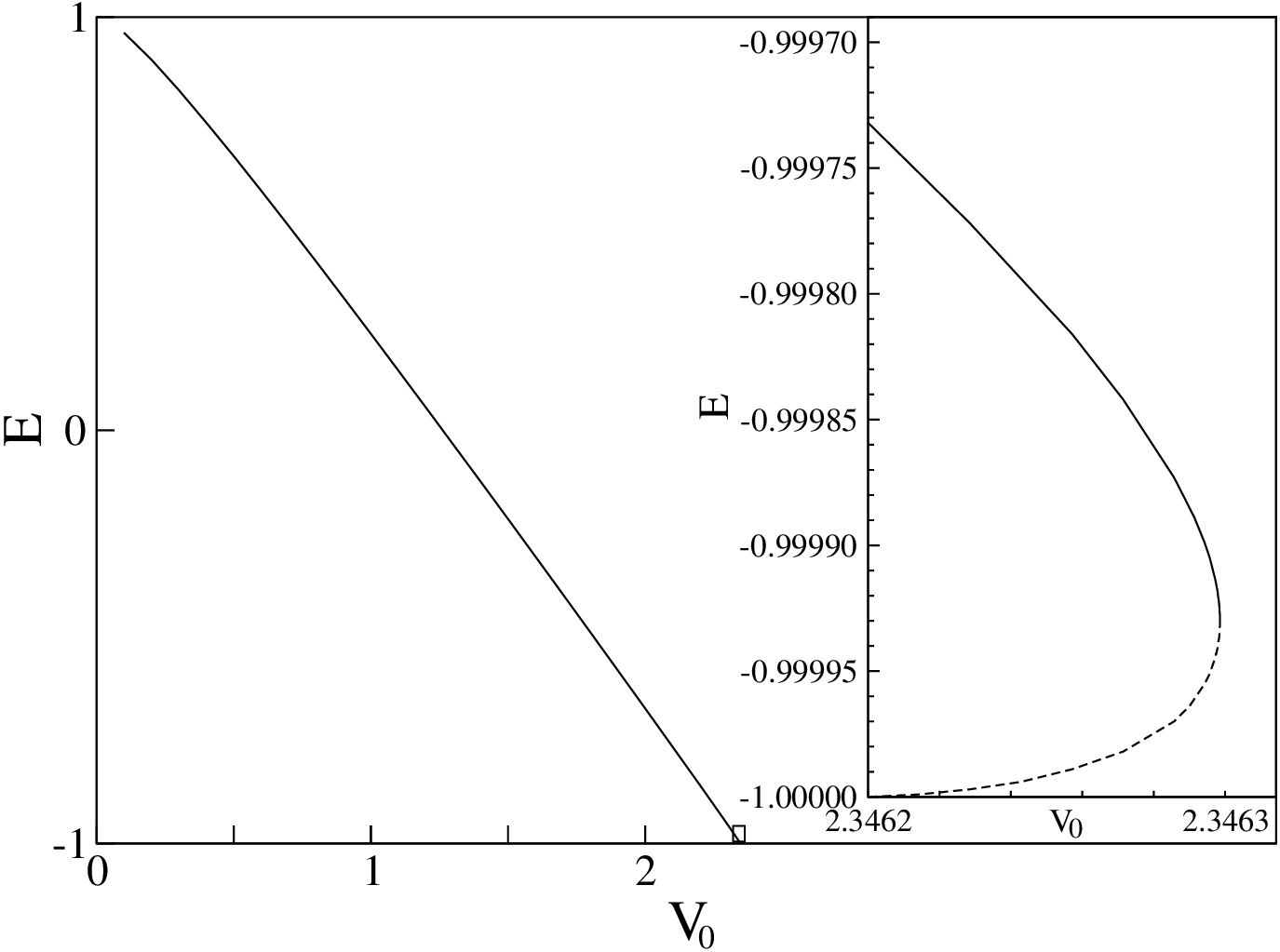,width=8.5cm}}
\vspace*{0pt} \caption{Energy of the lowest bound-state spectrum for
$L=2$, $a=2$. Inset is an enlargement of the critical area, showing
solid and dotted lines corresponding to positive and
negative norm state solutions respectively. The critical value for
$V_{0}=2.3463$ corresponds to $E=-0.999943$. Energy is given in units
of the rest energy $mc^2$} \label{fig:Fig3}
\end{figure}

\begin{figure}[th]
\vspace{1cm} \centerline{\psfig{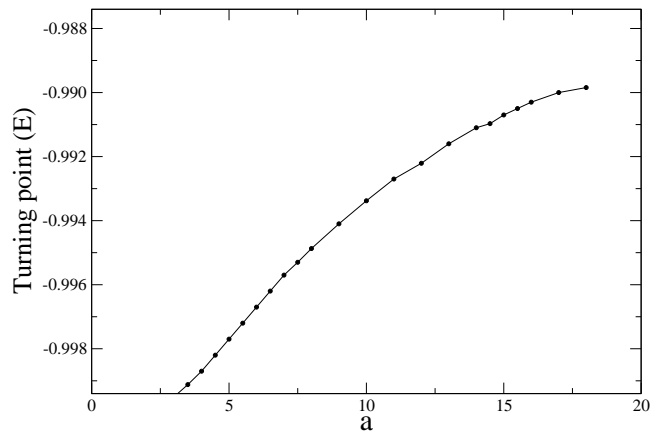}}
\vspace*{0pt} \caption{Turning point versus $a$ for $L=2$. Energy is
given in units of the rest energy $mc^2$} \label{fig:Fig4}
\end{figure}

\begin{figure}[th]
\vspace{1cm} \centerline{\psfig{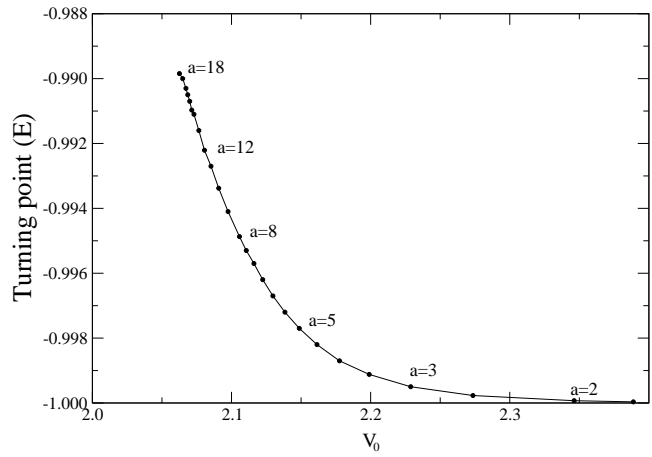}}
\vspace*{0pt} \caption{Turning point versus $V_0$ for $L=2$.}
\label{fig:Fig5}
\end{figure}

\section{The Cusp potential well}

\begin{figure}[th]
\vspace{1cm} \centerline{\psfig{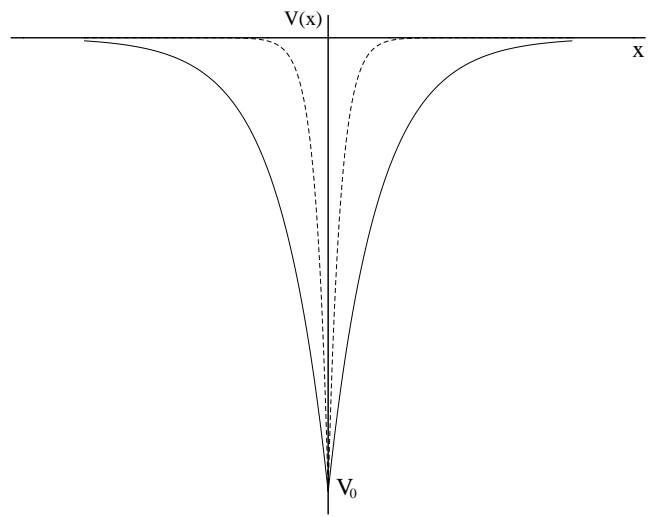}}
\vspace*{0pt} \caption{The Cusp potential well for $a=2$ (solid
line) and $a=0.5$ (dotted line).} \label{f1}
\end{figure}

In this section we are interested in studying the Klein-Gordon
equation (\ref {0}), with $p_{\perp }=0$, in the presence of a cusp
potential given by the expression \cite{Villalba}:

\begin{equation}  \label{1}
V(x)=\left\{
\begin{array}{cc}
-V_{0}e^{x/a} \quad \mathsf{for} \quad x<0, &  \\
-V_{0}e^{-x/a}\quad \mathsf{for} \quad x>0. &
\end{array}
\right.
\end{equation}
The form of the potential (\ref{1}) is shown Fig. (\ref{f1}). The parameter $%
V_{0}>0$ determines the depth of the cusp potential well. From Fig. (\ref{f1}%
) one readily notices that the potential becomes sharper as the shape parameter $%
a$ becomes smaller. In fact, the asymptotic limit of $V(x)$ as
$a\rightarrow 0$ is (-$V_{0}/a)\delta (x)$. The potential (\ref{1})
can also be regarded
as a limit case of the Woods-Saxon potential well as the shape parameter $%
a\gg 1$ \ and $L=0$ in Eq. (\ref{ws1}). As for the Woods-Saxon
potential well, we split the solutions into two regions.

Consider the bound states solution for $x<0$. We solve the
differential equation

\begin{equation}  \label{2}
\frac{d^{2}\phi _{L}(x)}{dx^{2}}+\left[ \left( E+V_{0}e^{x/a}\right) ^{2}-1%
\right] \phi _{L}(x)=0.
\end{equation}

On making the change of variables $y=2iaV_{0}e^{x/a}$, Eq. (\ref{2})
becomes

\begin{equation}
y\frac{d}{dy}\left( y\frac{\phi _{L}}{dy}\right) -\left[ \left(
iaE+y/2\right) ^{2}+a^{2}\right] \phi _{L}=0.  \label{3}
\end{equation}
Putting $\phi _{L}=y^{-1/2}f(y)$ we obtain the Whittaker
differential equation \cite{Abra}

\begin{equation}
f(y)^{\prime \prime }+\left[ -\frac{1}{4}-\frac{iaE}{y}+\frac{%
1/4-a^{2}(1-E^{2})}{y^{2}}\right] f(y)=0.  \label{4}
\end{equation}

The general solution of Eq. (\ref{4}) can be expressed in terms of
Whittaker functions \ $M_{\kappa ,\mu }(y)$ and $W_{\kappa ,\mu
}(y)$ as \cite{Abra}

\begin{equation}
\phi _{L}(y)=c_{1}y^{-1/2}M_{\kappa \mu }(y)+c_{2}y^{-1/2}W_{\kappa
\mu }(y), \label{5}
\end{equation}

\noindent where $c_{1}$ and $c_{2}$ are arbitrary constants, and
$\kappa $ and $\mu $ are

\begin{equation}
\kappa =-iaE,\hspace{0.3cm}\mu =a\sqrt{1-E^{2}}.  \label{6}
\end{equation}

Analogously, we proceed to obtain the solutions of Eq. (\ref{0}) in
the presence of the potential (\ref{1}) for $x>0.$ In this case we
have the differential equation

\begin{equation}
\frac{d^{2}\phi _{R}(x)}{dx^{2}}+\left[ \left( E+V_{0}e^{-x/a}\right) ^{2}-1%
\right] \phi _{R}(x)=0.  \label{7}
\end{equation}
On making the change of variables $z=2iaV_{0}e^{-x/a}$, Eq.
(\ref{7}) becomes

\begin{equation}
z\frac{d}{dz}\left( z\frac{\phi _{R}}{dz}\right) -\left[ \left(
iaE+z/2\right) ^{2}+a^{2}\right] \phi _{R}=0.  \label{8}
\end{equation}
Putting $\phi _{R}=z^{-1/2}g(z)$ we obtain the differential equation

\begin{equation}
g(z)^{\prime \prime }+\left[ -\frac{1}{4}-\frac{iaE}{z}+\frac{%
1/4-a^{2}(1-E^{2})}{z^{2}}\right] g(z)=0,  \label{9}
\end{equation}

\noindent whose general solution is

\begin{equation}
\phi _{R}(z)=d_{1}z^{-1/2}M_{\kappa \mu }(z)+d_{2}z^{-1/2}W_{\kappa
\mu }(z), \label{10}
\end{equation}
where $d_{1}$ and $d_{2}$ are arbitrary constants.

Since we are looking for bound states of Eq. (\ref{0}) with the potential (%
\ref{1}), we choose to work with regular solutions \ $\phi _{L}(y)$ and $%
\phi _{R}(z)$ along the $x$ axis:

\begin{equation}
\begin{array}{cc}
\phi _{L}(y)=c_{1}y^{-1/2}M_{\kappa ,\mu }(y), &  \\
\phi _{R}(z)=d_{1}z^{-1/2}M_{\kappa ,\mu }(z). &
\end{array}
\label{11}
\end{equation}

Imposing the condition that the left $\phi _{L}(y)$ and right $\phi
_{R}(z)$ scalar wave functions must be continuous at $x=0$ as well
as their first derivatives, we obtain that the energy eigenvalues
must satisfy the equation:

\begin{equation}
(1-2iaV_{0}+2\kappa )M_{\kappa \mu }(2iaV_{0})-(1+2\kappa +2\mu
)M_{{\kappa +1},\mu }(2iaV_{0})=0,  \label{12}
\end{equation}

\begin{figure}[th]
\vspace{1cm} \centerline{\psfig{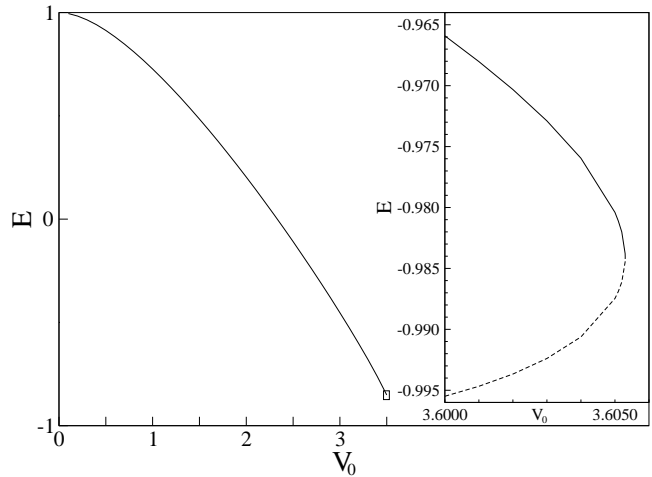}}
\vspace*{0pt} \caption{Energy of the lowest bound-state spectrum for
$a=0.5$. Inset is an enlargement of the critical area, showing
solid and dotted lines corresponding to positive and
negative norm state solutions respectively. The critical value for
$V_{0}=3.605$ corresponds to $E=-0.984766$. Energy is given in units of the
rest energy $mc^2$} \label{f2}
\end{figure}

\begin{figure}[th]
\vspace{1cm} \centerline{\psfig{file=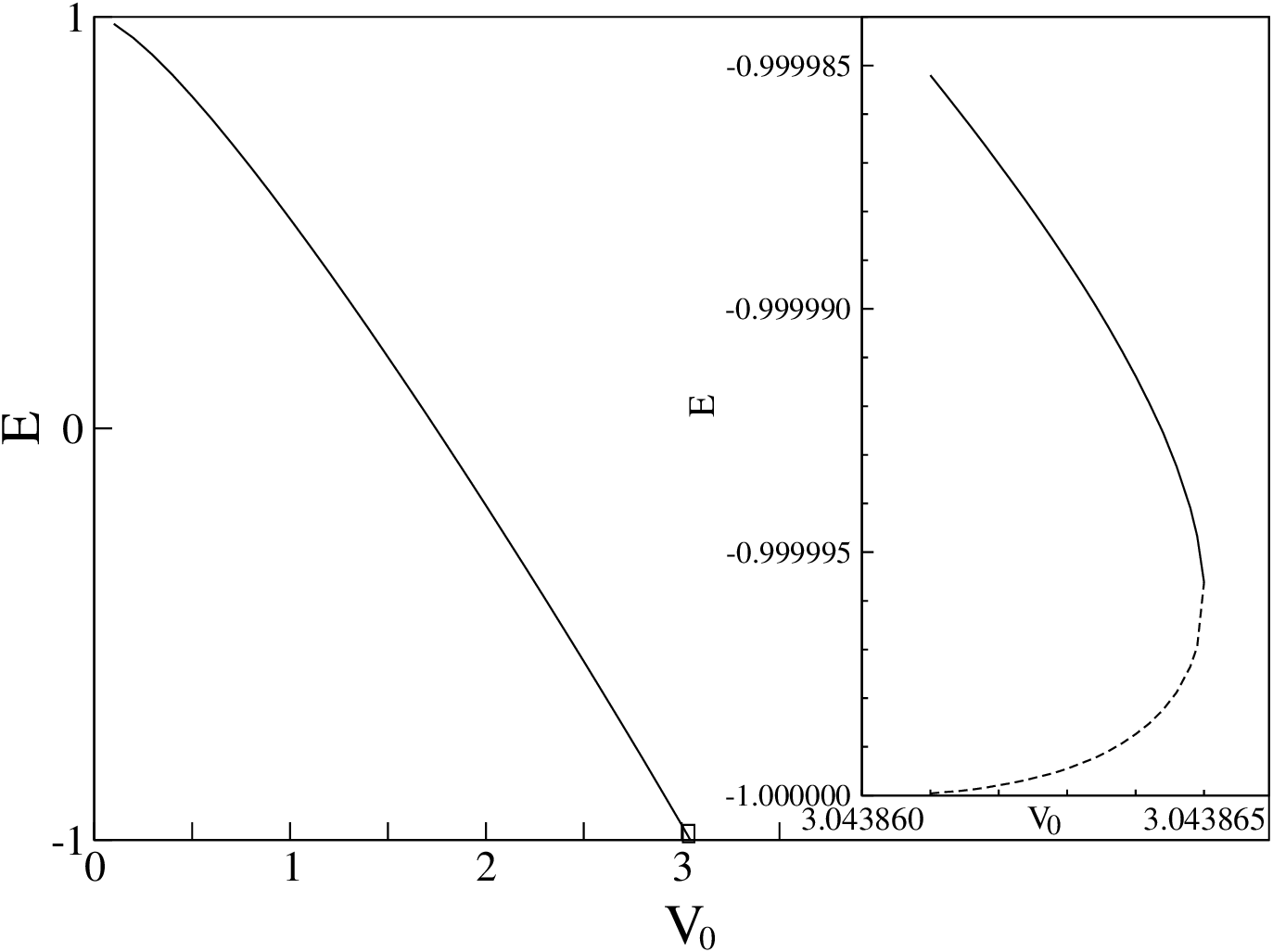,width=8.5cm}}
\vspace*{0pt} \caption{Energy of the lowest bound-state spectrum for
$a=1.2$. Inset is an enlargement of the critical area, showing
solid and dotted lines corresponding to positive and
negative norm state solutions respectively. The critical value for
$V_{0}=3.043865$ corresponds to $E=-0.999996$. Energy is given in units of the
rest energy $mc^2$} \label{f3}
\end{figure}

\begin{figure}[th]
\vspace{1cm} \centerline{\psfig{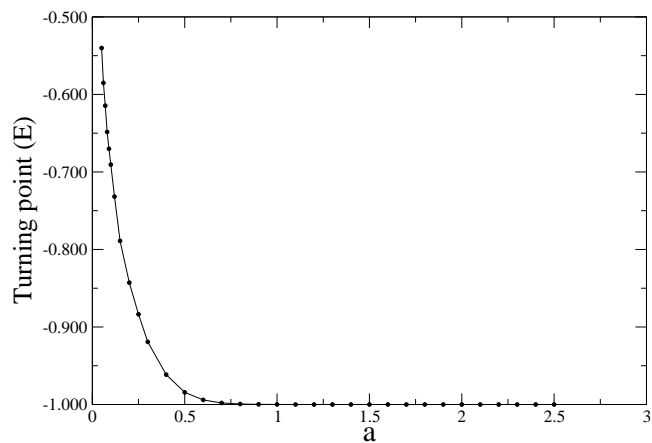}}
\vspace*{0pt} \caption{Turning point versus a. Energy is given in
units of the rest energy $mc^2$} \label{f4}
\end{figure}

\begin{figure}[th]
\vspace{1cm} \centerline{\psfig{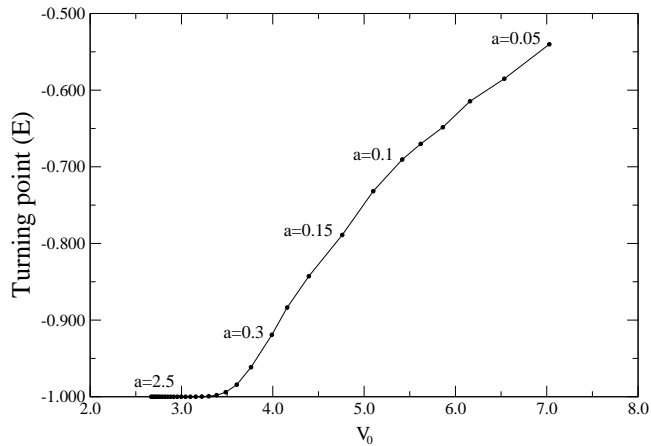}}
\vspace*{0pt} \caption{Turning point versus $V_0$. Energy is given
in units of the rest energy $mc^2$} \label{f5}
\end{figure}

The explicit solutions of Eq. (\ref{12}), showing the dependence of
the energy $E$ on $V_{0}$ and $a$ can be determined numerically. The
dependence of the ground state spectrum on the potential strength
$V_{0}$ is shown in Figs. (\ref{f2}) and (\ref{f3}). We can readily
see that for $a=0.5$, \ antiparticle states arise for $V>3.6050$
and, for $a=1.2$, \ antiparticle states arise for $V>3.04386$. Here
the solutions coming from the lower continuum into the region of
bound states meet with solutions from the upper continuum. This
leads to the possibility of spontaneous productions of pairs. As the
Woods-Saxon potential well, with the cusp potential well, the
particle bound states ($E^{(+)}$) and antiparticle bound states
($E^{(-)}$) correspond to $N>0$ and $N<0$ respectively. For $N=0$
both solutions meet and have the same energy. The normalization of
the wave functions (\ref{11}) is given by Eq. (\ref{norma}).
Antiparticle bound states arise for all values of the shape
parameter considered. We varied the shape parameter $a$ from $0.05$
to $2.5$. Figs. (\ref{f4}) and (\ref{f5}) show the behavior of the
turning point $(E)$ versus the potential parameters $a$ and $V_{0}$
respectively. Fig. (\ref{f4}) shows that as the value of $a$
increases, the energy value, for which antiparticle states appear,
decreases. In Fig. (\ref {f5}) we observe that, as the value of
$V_{0}$ increases, the energy value for which antiparticle states
appear increases.

\section{Conclusions}

In this article we have shown that the Woods-Saxon and the cusp
potential well are able to bind scalar particles.

The Woods-Saxon potential, analogous to the square well potential,
shows antiparticle bound states. The turning point, where the norm
is zero, depends on the potential parameters $a$ and $V_{0}$.
Therefore the one-dimensional Woods-Saxon potential exhibits a
behavior characteristic of short range potentials \cite{Popov}. It
should be expected that, like in the Coulomb problem,  for slow
damping potentials no antiparticle bound states appear. The results
reported in this article suggest that the one-dimensional
Schiff-Snyder-Weinberg \ effect \cite {Schiff} can be extended to
the case of potentials with non compact support, provided they
exhibit, for large values of the space parameter, a fast damping
asymptotic behavior.

\section*{Acknowledgments}

This work was supported by FONACIT under project G-2001000712.

\end{document}